\documentclass[preprint,amsmath,amssymb]{revtex4}
\usepackage{graphicx}
\usepackage{dcolumn}
\usepackage{bm}
\usepackage{pstricks}
\usepackage{epsfig}
\usepackage{pst-grad}
\usepackage{pst-plot}
\begin{document}
\title{The fermionic Unruh effect in relativistic eccentric uniformly
rotating frames}
\author{${\rm H.\;Ramezani}$-${\rm Aval}$\footnote{Electronic
address:~ramezani@gonabad.ac.ir}} \affiliation{ Department of
Physics, University of Gonabad, Gonabad, Iran}
\begin{abstract}
We investigate detection of Dirac quanta in a uniformly eccentric
rotating frame both with canonical and detector approaches by
employing relativistic rotational transformations. First we consider
a relativistic uniformly eccentric rotating detector that is coupled
to the scalar density of a massless Dirac field, and show that this
detector has a nonzero response function and observes a Planckian
energy spectrum. Then using relativistic rotational transformations,
we investigate the canonical quantization of Dirac field in
relativistic eccentric rotating frame. We show that it's not
possible to obtain the analytic solution for Dirac equation in this
frame and so canonical approach can not be carried out.
\end{abstract}
\maketitle
\section{Introduction} \label{sec1}
The Unruh effect predicts that particles will be detected in a
vacuum by an accelerated observer. These particles can be boson or
fermion and the observer may be an uniform accelerating observer
(Rindler observer) or a rotating observer. There are two approaches
to investigate Unruh effect in all of these cases. A nonlocal
approach to this effect is based on canonical quantization of field
in two different frames (Minkowski and accelerated frame) and then
calculating the particle number operator with the Bogoliubov
coefficients relating accelerating and Minkowski modes; nontrivial
Bogoliubov transformation between these modes concludes non-zero
expectation value of particle number operator of the accelerating
observer in the vacuum state of laboratory inertial observer. We
will refer to this approach as " the canonical approach". The second
was a local approach based on probing the field coupled to a
detector on finite patches of spacetime \cite{Davies}. Excitation of
detector is interpreted as detection of particle or antiparticle. We
will refer to latter
method as "detector approach".\\
 Most of the first researches on Unruh effect and quantum field theory in accelerated systems deal
with scalar fields and Rindler observer. Klein-Gordon equation was
often used and detection of bosons by the uniformly accelerating
observer in the vacuum of inertial observer was studied. So far also
most studies have used the simplest kind of Unruh-DeWitt detectors,
namely detectors that couple linearly to a bosonic real scalar
field. Although the physical interpretation is not entirely clear,
it is by now well established that the uniformly accelerated
detector (observer) observes a spectrum of Bose-Einstein particle in
a thermal bath of temperature $T = a/2\pi$, where \emph{a} is the
magnitude of the proper acceleration. Then similar methods was done
for spinor fields and detection of fermions by the Rindler observer
in the Minkowski vacuum was studied. First in \cite{Muller},
investigation of Unruh effect for fermion fields was done by
canonical approach. They analytically solve the Dirac equation in
the uniformly accelerating frame and then show that the Rindler
observer experiences a thermal flux of Dirac particles and
antiparticles with temperature $T = a/2\pi$ in the ordinary
Minkowski vacuum. This particle number distribution of the Minkowski
vacuum turned out to be thermal in the accelerated frame,
independently of the fermion mass. Zhu in \cite{Zhu} obtain the same
result by similar method, and  using canonical approach, show in
details the mathematical appearance of fermionic particle and
antiparticles in the Rindler frame. For a complete review on above three cases see \cite{Crispino}.\\
But for the fermionic Unruh effect, the detector approach has
different results. In \cite{Louko} for an Unruh-DeWitt detector that
is coupled linearly to the scalar density of a massless Dirac field,
it was shown that "in d-dimensional Minkowski the fermionic
detectors response is identical to that of a detector coupled
linearly to a massless scalar field in 2d-dimensional Minkowski. In
the special case of uniform linear acceleration, the detectors
response hence exhibits the Unruh effect with a Planckian factor in
both even and odd dimensions." On the other hand Takagi in
\cite{Takagi}, for the similar coupling between detector and
massless Dirac field, show that for massless Dirac field the Rindler
power spectrum is characterized by the Fermi distribution if the
dimension of spacetime is even and by the Bose
distribution if it is odd.\\
Unruh effect for rotating observer also has been investigated both
for boson and fermion fields. As we have mentioned in
\cite{Nouri1,Nouri2}, there are some ambiguities related to the
rotational phenomena. But this subject is interesting because
rotating observer has more feasible practical and experimental
advantages toward uniformly accelerating one. In \cite{Ramezani} we
reviewed the bosonic Unruh effect for rotating observer. In this
case, employing two relativistic rotational transformations
corresponding to the eccentric rotating observer, the agreement
between canonical approach and detector approach seems not to occur.
it was shown that in both cases, the detector response function is
nonzero but one of them has zero bogolioubov coefficient while the
other has nonzero. Those conclusions reinforce the claim that
correspondence between vacuum states defined via canonical field
theory and a detector is broken for bosonic fields in rotating
frames.\\
For the fermionic Unruh effect in rotating frames, the canonical
approach was first applied in \cite{Iyer}. Using the property that
for the Dirac field in contrast with scalar field all modes positive
and negative have a positive norm, and without explicit calculation
of Bogoliubov coefficient, Iyer shows that for rotating observer the
expectation value of number operator of particles and antiparticles
is nonzero in vacuum state of inertial Minkowski observer. Then in
\cite{Manning} an alternate quantization of fermions in rotating
reference frames is introduced and the Bogoliubov coefficients are
calculated, which give nonzero $\beta$ coefficient. \\
This history motivates us to develop the fermionic Unruh effect for
detector approach that has not yet been studied. On the other hand
we know that both rotating observer and fermionic fields are
experimentally more accessible than linearly accelerating observer
and bosonic fields. So it is important to study the fermionic unruh
effect in rotating frames. The aim of the present work is to study
this important case in a uniformly eccentric rotating frame both
with canonical and detector approach by
employing relativistic rotational transformations.\\
In section II we introduce the relativistic eccentric uniformly
rotating frame, which we use in this paper. In section III we obtain
the response of relativistic eccentric rotating detector in the
fermionic vacuum state of inertial observer. In section IV, using
relativistic rotational transformations, we investigate the
canonical quantization of Dirac field in relativistic eccentric
rotating frame.\\
We use natural units in which $\hbar=c=1$ and the metric with
signature $(-,+,+,+)$.
\section{Relativistic eccentric uniformly rotating frame} \label{sec4}
\begin{figure}[h]
    \includegraphics{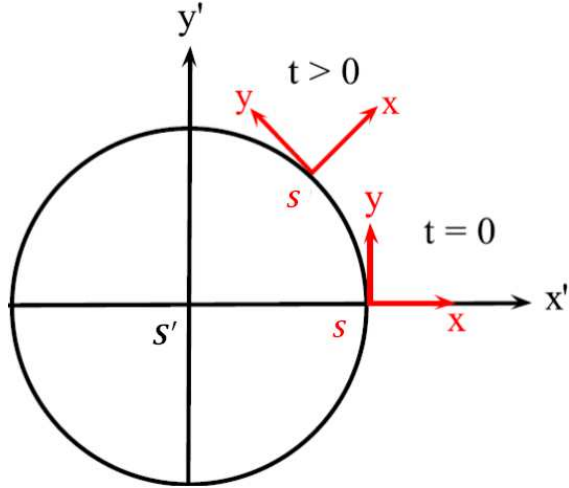}
    \caption{The detector position is in the origin of rotating frame.}
    \label{position}
\end{figure}

Study of physical phenomenon in rotating frames always encounters
some ambiguity. The most important of them is finding a set of
transformations that can give the correct relation between rotating
and inertial observers. The Galilean transformations are usually
used in these works, including \cite{Iyer, Manning}, but as we
mentioned in \cite{Nouri1, Nouri2}, using Galilean rotational
transformations for an eccentric rotating detector is not true,
while the real experimental setups is the case of
observers/detectors on the circumference of a uniformly rotating
disk. So it is necessary to consider relativistic properties and
eccentric position of a rotating observer.\\
In \cite{MTW} based on consecutive Lorentz transformations and Fermi
coordinates, the general form of the metric in an accelerated
spinning frame in curved spacetime is derived as follows:
\begin{eqnarray}\label{5}
ds^2=-{dx^{0}}^2[1+2a_{j}x^{j}+(a_{l}x^{l})^2+(\Omega_{l}x^{l})^2-\Omega^{2}x_{l}x^{l}+R_{0l0m}x^{l}x^{m}]\nonumber
\\
+2dx^{0}dx^{i}(\epsilon_{ijk}\Omega^{j}x^{k}-\frac{2}{3}R_{0lim}x^{l}x^{m})+dx^{i}dx^{j}(\delta_{ij}-\frac{1}{3}R_{iljm}x^{l}x^{m})
\end{eqnarray}
which in flat spacetime reduces to
\begin{eqnarray}\label{6}
ds^2=-{dx^{0}}^2[1+2a_{l}x^{l}+(a_{l}x^{l})^2+(\Omega_{l}x^{l})^2-\Omega^2x_{l}x^{l}]\nonumber
\\
+2dx^{0}dx^{i}(\epsilon_{ijk}\Omega^{j}x^{k})+dx^{i}dx^{j}\delta_{ij}.
\end{eqnarray}
in which ${\bf a}$ and ${\bf \Omega}$ are the observer's
acceleration and (spin) rotation measured in a comoving inertial
frame $(S)$ whose velocity is momentarily the same as that of the
accelerating observer. In our setup the origin of the rotating frame
is on the rim of the circular path as shown in Figure \ref{position}
(See \cite{Nouri2} for more details.).
So as given in \cite{Mashhoon} the component
of detector's acceleration and (spin) rotation measured in a
comoving inertial frame are
\begin{eqnarray}\label{11}
{a}^l=(-\gamma^{2}R\Omega^2,0,0) ~~~,~~~
{\Omega}^l=(0,0,\gamma^{2}\Omega)
\end{eqnarray}
On the other hand, In \cite{Mashhoon, Nikolic} Special Relativistic
coordinate Transformation (SRT)between inertial laboratory (primed)
and eccentric uniformly rotating (unprimed) frames are given as
follows:
\begin{eqnarray}\label{2}
t'=\gamma(t+R\Omega y)~~~x'=(x+R)\cos(\gamma\Omega t)-\gamma
y\sin(\gamma\Omega t) \nonumber
\\
y'=\gamma y\cos(\gamma\Omega t)+(x+R)\sin(\gamma\Omega
t)~~~~,~~~~z=z^{\prime}
\end{eqnarray}
in which $\gamma=(1-R^2\Omega^2)^{-1/2}$, $\Omega$ is the uniform
angular velocity of the disk and R is the radius of the circular
path. We will use these results in the next sections.
\section{Response of relativistic eccentric rotating detector to Dirac field} \label{sec4}
\begin{figure}[h]
    \includegraphics[height = 6cm, width=5cm]{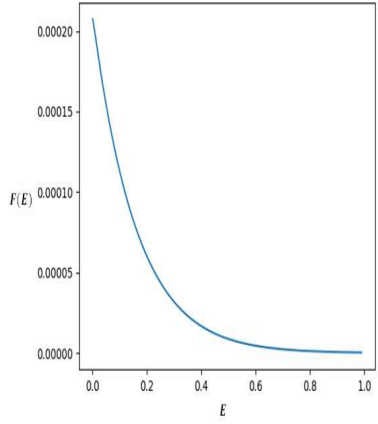}
    \caption{The detector response function of fermionic field.}
    \label{int}
\end{figure}

\begin{figure}[h]
    \includegraphics[height = 6cm, width=5cm]{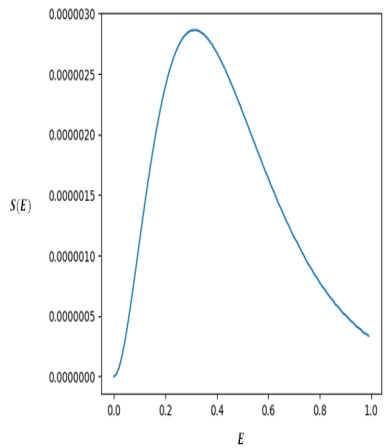}
    \caption{Energy spectrum of relativistic rotating detector in fermionic field.}
    \label{E2int}
\end{figure}

In this section we use detector approach to investigate fermionic
Unruh effect. We consider a pointlike detector model that has been
commonly employed in the literature (for example, see
\cite{Sriramkumar} and references  therein) and parametrize the
detector's trajectory by it's proper time $\tau$ and so it moves on
the worldline $x(\tau)$ and then allowing the detector to be
switched on and off in an arbitrary smooth way\cite{Hummer}. We know
that as mentioned in \cite{Hummer}, any interaction Hamiltonian
between an Unruh Dewitt detector and a field must be a self-adjoint
scalar. In the case of quantized spinor field, the simplest
self-adjoint Lorentz scalar is quadratic term $\overline{\psi}\psi$.
Furthermore, an Unruh Dewitt detector coupled linearly to a fermion
field would be able to violate fermion number conservation by
creating and annihilating individual fermions. So this detector is a
two level system coupled to the scalar density of a massless Dirac
field by the interaction Hamiltonian
\begin{eqnarray}\label{10}
H_{int}:=cx(\tau)m(\tau)\overline{\psi}(x(\tau))\psi(x(\tau))
\end{eqnarray}
where $m(\tau)$ is the detector's monopole moment operator and
$x(\tau)$ is switching function; a smooth real-valued function that
specifies how the interaction is turned on and off. Corresponding to
this interaction the detector response function is given by
\cite{Davies}
\begin{eqnarray}\label{10}
\emph{F}=\int{d\tau d\tau'
x(\tau)x(\tau')e^{-iE(\tau-\tau')}W^{(2,\overline{2})}}
\end{eqnarray}
where
\begin{eqnarray}\label{10}
W^{(2,\overline{2})}=\langle \psi_0 |
\overline{\psi}(x)\psi(x)\overline{\psi}(y)\psi(y)|\psi_0\rangle
\end{eqnarray}
is the correlation function. For our present purpose, we use the
theorem 1 in \cite{Louko}: The response function of an Unruh-DeWitt
detector coupled quadratically to a massless Dirac field in
Minkowski vacuum in $d\geq2$ spacetime dimensions,$~F(d)$, equals
\begin{eqnarray}\label{Louko Theorem}
\frac{N_d(\Gamma(d/2))^2}{\Gamma(d-1)}
\end{eqnarray}
times the response function of an Unruh-DeWitt detector coupled
linearly to a massless scalar field in Minkowski vacuum in 2d
spacetime dimensions,$~F_s(2d)$. In (\ref{Louko Theorem})
\begin{eqnarray}\label{10}
N_d=\left\{
      \begin{array}{ll}
        2^{d/2}, & \hbox{for d even;} \\
        2^{(d-1)/2}, & \hbox{for d odd.}
      \end{array}
    \right.
\end{eqnarray}
 So
\begin{eqnarray}\label{10}
F(4)=\frac{4(\Gamma(2))^2}{\Gamma(3)}F_s(8)=2F_s(8)
\end{eqnarray}
 So we need the response function
of Unruh-DeWitt detector that is linearly coupled to a scalar field,
which is given by
\begin{eqnarray}\label{12}
\emph{F}_s=\int_{-\infty}^{\infty}d\triangle \tau
e^{-iE\triangle\tau}G^{+}(x(\tau_1),x(\tau_2))
\end{eqnarray}
where $G^{+}$ is the scalar fields positive Wightman function. In
the case of a massless scalar field for $d>2$ we have
\cite{Décanini}
\begin{eqnarray}\label{10}
G^{+}(x,y)=\frac{\Gamma(d/2-1)}{4\pi^{d/2}(z(x,y))^{d-2}}
\end{eqnarray}
in which
\begin{eqnarray}\label{10}
z=[(\textbf{x-y})^2-(x^0-y^0-i\epsilon)^2]^{1/2}
\end{eqnarray}
and so for $d=8$ we have
\begin{eqnarray}\label{10}
G^{+}(x,y)=\frac{1}{2\pi^4(z(x,y))^{6}}
\end{eqnarray}
With the assumption that the detector is at the origin of rotating
frame $(x=0,y=0)$ and using transformations (\ref{2}), the
detector's trajectory in the laboratory frame is given by
\begin{eqnarray}\label{21}
x'=R \cos(\gamma\Omega t)~~~,~~~y'=R \sin(\gamma\Omega
t)~~~,~~~z'=0~~~,~~~t'=\gamma t
\end{eqnarray}
 and finally in this case we have
\begin{eqnarray}\label{22}
G^{+}(x'_1,x'_2)=\frac{1}{\pi^4[(\gamma\triangle\tau-i\epsilon)^2-2R^2(1-\cos(\gamma\Omega
\triangle\tau))]^3}
\end{eqnarray}
Inserting (\ref{22}) in (\ref{12}) the detector response function is
given by
\begin{eqnarray}\label{23}
\emph{F}(E)=\int_{-\infty}^{\infty}d\triangle \tau
\frac{e^{-iE\triangle\tau}}{[(\gamma\triangle\tau-i\epsilon)^2-4R^2\sin^2(\frac{\gamma\Omega
\triangle\tau}{2})]^3}.
\end{eqnarray}
for a numerical calculation we set $R=1$ and $\Omega=\frac{1}{2}$.
We following the procedure of Letaw in \cite{Letaw, Gutti}. In ordre
to eliminate the singularity in the integrand, we subtract the the
expansion of the zero point
  \begin{eqnarray}\label{24}
 series(\dfrac{1}{(\dfrac{4}{3}{\triangle\tau}^2-4\sin(\dfrac{1}{6}{\triangle\tau}\sqrt{3})^{2})^3})={\triangle\tau}^{-6}-\dfrac{1}{36}{\triangle\tau}^{-4}+\dfrac{1}{1215}{\triangle\tau}^{-2}+O({\triangle\tau}^{0})
  \end{eqnarray}
and numerically compute the integral. We obtain the nonzero detector
response function as a function of energy, as shown in Figure
\ref{int}.
and according to \cite{Letaw} we express the energy spectrum as a
function of energy is
\begin{eqnarray}\label{23}
S(E)=\frac{E^2}{4\pi^3}\emph{F}(E)
\end{eqnarray}
Again, in Figure \ref{E2int} we have plotted the energy spectrum as
a function of energy.
So the energy spectrum has a Planckian form. This result is in
agreement with Theorem 1 in \cite{Louko} that implies that the
response function of the Dirac field detector involves a Planck
factor in all spacetime dimensions.
\section{Canonical Quantization of Dirac field in eccentric uniformly rotating frame } \label{sec3}
 As we mention in previous section, if we use canonical approach we
need canonical quantization of field in rotating frame. We must
solve analytically the dirac equation in rotating frame and then
calculate the particle number operator with the Bogoliubov
coefficients relating rotating and Minkowski modes.\\
 In \cite{Hehl} the Dirac equation in the observer's local frame
according to \cite{Bjorken} is given by
\begin{eqnarray}\label{Dirac Equation}
i\hbar\gamma^{\mu}D_{\mu}\Psi=mc\Psi
\end{eqnarray}
in which
\begin{eqnarray}\label{7}
D_{0}=\frac{1}{1+\textbf{a}\cdot
\textbf{X}}[\frac{\partial}{\partial
x^{0}}+\frac{1}{2c^2}\textbf{a}\cdot\mathbf{\alpha}-\frac{i}{c
\hbar}\mathbf{\omega}\cdot\textbf{J}]~~~,~~~D_{i}=\frac{\partial}{\partial
x^{i}}
\end{eqnarray}
where
\begin{eqnarray}\label{11}
\textbf{J=L+S=X}\times \frac{\hbar}{i}\frac{\partial}{\partial
\textbf{X}}+\frac{1}{2}\hbar \sigma
\end{eqnarray}
is the total angular momentum. The $\alpha$ matrices are denoted as
\begin{eqnarray}\label{12}
\alpha^{i}=\gamma^{0}\gamma^{i}
\end{eqnarray}
and $\sigma^{i}$ correspond to the three Pauli matrices. Because of
regularization problem related to massive Dirac field
\cite{Langlois}, here we focus on the massless field.\\

We use the following standard representation for Dirac matrices
\begin{eqnarray}\label{12}
\gamma^{0}=\left(
             \begin{array}{cc}
               1 & 0 \\
               0 & -1 \\
             \end{array}
           \right)
           ~~~~~ , ~~~~
           \gamma^{i}=\left(
             \begin{array}{cc}
               0 & \sigma_i \\
               -\sigma_i & 0 \\
             \end{array}
           \right)
\end{eqnarray},
so we have
\begin{eqnarray}\label{}
\textbf{a}\cdot\mathbf{\alpha}=a^1 \alpha^1=-\gamma^2 R \Omega^2
\alpha^1 \\
\mathbf{\omega}\cdot \textbf{J} =\gamma^2 \Omega J_z =\gamma^2\Omega
(\frac{\hbar}{i}x \frac{\partial}{\partial y}-\frac{\hbar}{i}y
\frac{\partial}{\partial x})+\frac{1}{2}\gamma^2\Omega\hbar \sigma^3
\end{eqnarray}
 And we denote Dirac field by
\begin{eqnarray}\label{10}
\psi=(\eta_1,\eta_2)^T
\end{eqnarray}
where $\eta_1$ and $\eta_2$ are two component spinors. As mentioned
in \cite{Iyer}, if Dirac field is massless then we have:
\begin{eqnarray}\label{10}
\eta_1=\eta_2\equiv(R(t,X),F(t,X))^T
\end{eqnarray}
Finally by inserting all of them in (\ref{Dirac Equation}) we obtain
two independent partial differential equations as below
\begin{eqnarray}\label{10}
\frac{1}{c^2-\gamma^2r\Omega^2 x}[c^2 \frac{\partial R}{\partial
t}-\gamma^2 c \Omega x \frac{\partial R}{\partial y}+\gamma^2 c
\Omega y \frac{\partial R}{\partial x}-\frac{i}{2}\gamma^2 c \Omega
R-\frac{1}{2}\gamma^2 r \Omega^2 F]+\frac{\partial R}{\partial
z}+\frac{\partial F}{\partial x}-i\frac{\partial F}{\partial
y}=0\nonumber \\
 \frac{1}{c^2-\gamma^2r\Omega^2 x}[c^2 \frac{\partial
F}{\partial t}-\gamma^2 c \Omega x \frac{\partial F}{\partial
y}+\gamma^2 c \Omega y \frac{\partial F}{\partial
x}+\frac{i}{2}\gamma^2 c \Omega F-\frac{1}{2}\gamma^2 r \Omega^2
R]-\frac{\partial F}{\partial z}+\frac{\partial R}{\partial
x}+i\frac{\partial R}{\partial y}=0\nonumber \\
\end{eqnarray}
because all terms in these equations are independent of z and t
components, we can choose R and F functions as
\begin{eqnarray}\label{10}
R=e^{ikz+i\omega t}R(x,y)~~~,~~~F=e^{ikz+i\omega t}F(x,y)
\end{eqnarray}
It's necessary to obtain the analytic solution for this partial
differential equation if we want to calculate Bogoliubov coefficient
and obtain expectation value of particle number operator of the
rotating observer in the vacuum state of laboratory observer.
However, unfortunately, it does not seem to be possible to evaluate
the corresponding solution analytically. So Because of computational
problems, it seems that this approach can not give a final
conclusion for fermionic Unruh effect in relativistic eccentric
rotating observer.
\section{Conclusion} \label{sec5}
We investigated the fermionic Unruh effect in a uniformly eccentric
rotating frame both with canonical and detector approach by
employing relativistic rotational transformations.  We show that
uniformly eccentric rotating detector that is coupled to the scalar
density of a massless Dirac field effectively immersed in a bath of
particles with a Planckian energy spectrum. This result is in
agreement with previous result that implies that the response
function of the Dirac field detector involves a Planck factor in all
spacetime dimensions. Then using relativistic rotational
transformations, we investigated the canonical quantization of Dirac
field in relativistic eccentric rotating frame. We showed that it's
not possible to obtain the analytic solution for Dirac equation in
this frame and so canonical approach
can not be carried out.\\
\\
\textbf{Acknowledgments}\\
I am grateful to M. Ansari-Rad and V. Anari for helpful discussions
on numerical calculation. This work was supported by the University
of Gonabad.

\end{document}